\documentclass[reprint,superscriptaddress,nofootinbib,amsmath,amssymb,aps,pra]{revtex4-2}

\usepackage{graphicx}
\usepackage{dcolumn}
\usepackage{bm}
\usepackage{braket}
\usepackage{mathrsfs}
\usepackage{hyperref}
\hypersetup{
    colorlinks,%
    citecolor=blue,%
    filecolor=blue,%
    linkcolor=blue,%
   	urlcolor=blue,
   	linktoc=page
}
\allowdisplaybreaks
\usepackage[framemethod=tikz]{mdframed}

\newcommand{\D}{\text{d}}

\usepackage{tikz}
\usetikzlibrary{calc}
\usetikzlibrary{decorations.pathreplacing}
\usetikzlibrary{decorations.pathmorphing}
\usetikzlibrary{decorations.markings}
\usetikzlibrary{arrows.meta}
\usetikzlibrary{positioning}

\begin{document}

\preprint{APS/123-QED}

\title{Superrotations at Spacelike Infinity}

\author{Adrien Fiorucci}
\email{adrien.fiorucci@tuwien.ac.at}
\affiliation{%
Institute for Theoretical Physics, TU Wien\\
Wiedner Hauptstrasse 8–10/136, A-1040 Vienna, Austria
}%

\author{Javier Matulich}
\email{javier.matulich@csic.es}
\affiliation{%
Instituto de Física Teórica UAM/CSIC, Calle Nicolás Cabrera 13-15, Universidad Autónoma de Madrid, Cantoblanco, Madrid 28049, Spain.}

\author{Romain Ruzziconi}
\email{romain.ruzziconi@maths.ox.ac.uk}
\affiliation{%
Mathematical Institute, University of Oxford, \\ Andrew Wiles Building, Radcliffe Observatory Quarter, \\
Woodstock Road, Oxford, OX2 6GG, UK
}

\date{\today}

\begin{abstract}
We propose a consistent set of boundary conditions for gravity in asymptotically flat spacetime at spacelike infinity, which yields an enhancement of the Bondi-Metzner-Sachs group with smooth superrotations and new subleading symmetries. These boundary conditions are obtained by allowing fluctuations of the boundary structure which are responsible for divergences in the symplectic form, and a renormalization procedure is required to obtain finite canonical generators. The latter are then made integrable by incorporating boundary terms into the symplectic structure, which naturally derive from a linearized spin-two boundary field on a curved background with positive cosmological constant. Finally, we show that the canonical generators form a nonlinear algebra under the Poisson bracket and verify the consistency of this structure with the Jacobi identity.
\end{abstract}

\maketitle

\section{Introduction}

Gravity in four-dimensional asymptotically flat spacetime constitutes a model for a large range of phenomena, from the scattering of elementary particles to the description of astrophysical systems below the cosmological scale. It came as a surprise when Bondi, van der Burg, Metzner and Sachs highlighted that the asymptotic symmetry group of such spacetimes is not only Poincaré, but an infinite-dimensional enhancement of the latter, called the (global) BMS group \cite{Bondi:1962px,Sachs:1962zza}, which is a semi-direct product between the Lorentz group and the supertranslations.

It was later shown that this asymptotic symmetry group could itself be consistently enhanced by including all the conformal transformations on the two-dimensional celestial sphere, called the superrotations \cite{Barnich:2009se,Barnich:2010ojg,Barnich:2010eb,Barnich:2011mi} (see also \cite{Barnich:2013axa,Flanagan:2015pxa,Barnich:2019vzx,Barnich:2021dta}). Further extensions were obtained by considering asymptotically locally flat spacetimes with fluctuating boundary metric at null infinity, allowing for all the diffeomorphisms on the celestial sphere \cite{Campiglia:2014yka,Campiglia:2015yka,Compere:2018ylh,Flanagan:2019vbl,Campiglia:2020qvc}, or Weyl symmetries \cite{Freidel:2021fxf,Freidel:2021cjp}, see \textit{e.g.} \cite{Ruzziconi:2020cjt,Fiorucci:2021pha} for recent reviews. 

The BMS symmetries play a major role in the understanding of the infrared structure of quantum gravity \cite{Strominger:2017zoo}. Remarkably, assuming antipodal matching conditions between past and future null infinity, it was shown that the supertranslation Ward identity is equivalent to Weinberg's leading soft graviton theorem \cite{Strominger:2013jfa,He:2014laa}. Similarly, superrotation invariance of the gravitational scattering was shown to be related to the subleading soft graviton theorem \cite{Cachazo:2014fwa,Kapec:2014opa,Adamo:2014yya} at all orders of perturbation \cite{Bern:2014oka,He:2017fsb,Laddha:2018myi,Sahoo:2018lxl,Donnay:2021wrk,Donnay:2022hkf,Pasterski:2022djr,Agrawal:2023zea,Choi:2024ygx}. The BMS transformations were also explicitly connected to memory effects, which constitute potential observable phenomena in gravitational astronomy (see \textit{e.g.} \cite{Grant:2022bla}). More precisely, the supertranslations describe a displacement memory effect \cite{Strominger:2014pwa}, while the superrotations correspond to spin \cite{Pasterski:2015tva}, center-of-mass \cite{Nichols:2018qac} and superboost/velocity kick memories \cite{Compere:2018ylh}. 

The global BMS symmetries were originally discovered by studying the boundary structure of asymptotically flat spacetimes at null infinity. Strikingly, while the Poincaré symmetries had been found for a long time at spacelike infinity \cite{Regge:1974zd}, it is only recently that the BMS group has been uncovered there \cite{Troessaert:2017jcm,Henneaux:2018cst,Henneaux:2018hdj,Henneaux:2019yax}. The advantage of working at spacelike infinity is that it requires less regularity assumptions on the class of spacetimes that are considered, since it does not assume the existence of a smooth null infinity \cite{Friedrich:2017cjg,Kehrberger:2021uvf,Kehrberger:2021vhp,Valiente-Kroon:2002xys,Chrusciel:1993hx,Chrusciel:2002vb}. Furthermore, as discussed in \cite{Troessaert:2017jcm,Compere:2017knf,Henneaux:2018cst, Henneaux:2018hdj,Capone:2022gme,Compere:2023qoa,Prabhu:2019fsp,Prabhu:2021cgk, Mohamed:2023jwv}, it allows to derive the antipodal matching conditions necessary to establish the equivalence between BMS Ward identities and soft theorems \cite{Strominger:2013jfa,He:2014laa,Kapec:2014opa} in a well-posed formulation of the scattering problem. 

However, up to now, only the global version of the BMS group has been found at spacelike infinity. In particular, the superrotations are still missing, and due to their crucial role for the physics in asymptotically flat spacetime, it is of paramount importance to find these symmetries at spacelike infinity. In this paper, we address this challenging problem through a purely Hamiltonian approach and derive a phase space at spacelike infinity that accommodates smooth superrotations, coined as \textit{Spi superrotations} (this terminology is introduced by analogy with the Spi supetranslations of \cite{BeigIntegration,BeigSchmidt,Ashtekar:1991vb}). To do so, inspired by the analysis at null infinity \cite{Campiglia:2014yka,Campiglia:2015yka,Compere:2018ylh}, we propose a new set of boundary conditions for asymptotically locally flat spacetimes by relaxing those considered in \cite{Troessaert:2017jcm,Henneaux:2018cst,Henneaux:2018hdj,Henneaux:2019yax} and allowing some fluctuation of the boundary structure. This makes the analysis technically very demanding and requires the use of recently developed methods: $(i)$ a renormalisation of the action and the symplectic form is necessary to ensure the finiteness of the canonical generators (see \textit{e.g.} \cite{deHaro:2000vlm,Papadimitriou:2005ii,Mann:2005yr,Compere:2008us,Compere:2018ylh,Fiorucci:2020xto,Compere:2020lrt,Ruzziconi:2020wrb,Chandrasekaran:2021vyu,Freidel:2021fxf,Freidel:2021cjp,Grumiller:2021cwg,Geiller:2021vpg,Geiller:2022vto,Campoleoni:2022wmf,McNees:2023tus,Capone:2023roc,Freidel:2024tpl}); $(ii)$ a field-dependent redefinition of the symmetry parameters \cite{Barnich:2007bf,Compere:2017knf,Grumiller:2019fmp,Adami:2020ugu,Alessio:2020ioh,Ruzziconi:2020wrb,Adami:2020amw,Adami:2021nnf,Grumiller:2023ahv}, as well as the introduction of boundary degrees of freedom \cite{Henneaux:2018gfi, Fuentealba:2020aax, Fuentealba:2021xhn}, are needed to render the canonical generators integrable; $(iii)$ the treatment of non-linear terms to compute the resulting asymptotic symmetry algebra, which is reminiscent of what happens in higher dimensions where already the global BMS group is realised as a non-linear algebra at spacelike infinity \cite{Fuentealba:2022yqt,Fuentealba:2021yvo}.

The core of the paper is supplemented by an \hyperref[sec:Appendix]{Appendix} collecting technical and intermediate results that are not essential to follow the reasoning but shall be quoted for completeness. 
\section{Solution space}

The canonical action for Einstein's gravity in four dimensions is given by \cite{Dirac:1958sc,Arnowitt:1962hi,Hanson:1976cn}
\begin{equation}
    \mathcal S = \int \D t\,\D^3 x\,\mathcal L_H,\quad \mathcal L_H = \pi^{ij}\dot g_{ij} - N\mathcal H - N^i\mathcal H_i \label{action}
\end{equation}
where $g_{ij}$ is the three-dimensional metric on the (spacelike) constant time  slices, $\pi^{ij}$ its conjugate momentum and $N$, $N^i$ correspond respectively to the lapse and shift functions. We work in units such that $16\pi G = 1$. We write $R_{ij}$ the spatial Ricci curvature, $R$ its trace, $_{\vert i}$ the spatial covariant derivative and $g = \det g_{ij}$. The variation of the action with respect to the lapse and the shift yields the Hamiltonian and momentum constraints 
\begin{equation}\label{constraints}
\begin{split}
    \mathcal H  &= - \sqrt g R + \frac{1}{\sqrt g} \big(\pi^{ij} \pi_{ij} -
	\frac{1}{2} \pi^2\big)\approx 0, \\
    \mathcal H_i  &= -2 {{\pi_i}^j}_{|j}\approx 0. 
\end{split}
\end{equation}
In the following of the paper, the weak equality symbol $\approx$ denotes the equality on the constraint surface. We introduce spherical coordinates $x^i = (r,x^A)$ on constant time slices and we are interested in the asymptotic behaviour of the dynamical fields around spacelike infinity, located on the two sphere of radius $r\to+\infty$. We propose the following boundary conditions: 
\begin{equation}\label{BC}
\begin{split}
    g_{rr} &= 1+\frac{\bar{h}_{rr}}{r}+\frac{\bar{h}^{(2)}_{rr}}{r^2}+ \mathcal O(r^{-3}), \\
    g_{rA} &= \bar{\lambda}_{A} + \frac{\bar{h}_{rA}}{r}+\mathcal O(r^{-2}), \\
    g_{AB} &= r^2 \bar{G}_{AB} + r \bar{h}_{AB} + \bar{h}^{(2)}_{AB}+\mathcal O(r^{-1}) , \\
    \pi^{rr} &= r \bar{P}^{rr}+\bar{\pi}^{rr}+\frac{\bar{\pi}^{rr}_{(2)}}{r}+ \mathcal O(r^{-2}), \\
    \pi^{rA} &=  \frac{\bar{\pi}^{rA}}{r}+ \frac{\bar{\pi}^{rA}_{(2)}}{r^2}+\mathcal O(r^{-3}), \\
    \pi^{AB} &=  \frac{\bar{P}^{AB}}{r} +\frac{\bar{\pi}^{AB}}{r^2}+\frac{\bar{\pi}_{(2)}^{AB}}{r^3} +\mathcal O(r^{-4}),
\end{split}
\end{equation}
where the transverse boundary metric $\bar G_{AB}$ is a general two-dimensional metric on the sphere. 
We also impose the following parity conditions:
\begin{equation}\label{parity}
    \begin{split}
        &\bar G_{AB} \sim \bar h_{rr} \sim \text{even},\quad \bar P^{AB}\sim \text{odd}, \\
        &\bar\pi^{rr} - \bar \pi + \bar h_{rr}\bar P - \bar P^{AB}\bar h_{AB} \sim\text{odd}, 
    \end{split}
\end{equation}
under the antipodal map $x^A\mapsto -x^A$ on the sphere. Throughout, we denote the trace of any tensor $T_{AB}$ on the sphere by $T = \bar G^{AB} T_{AB}$.

These boundary conditions are weaker than those considered in \cite{Troessaert:2017jcm,Henneaux:2018cst,Henneaux:2018hdj,Henneaux:2019yax} since now, the leading orders $\bar{G}_{AB}$, $\bar{P}_{AB}$ and $\bar{P}^{rr}$ are allowed to fluctuate (by this, we mean that they are not fixed on the solution space). One recovers the boundary conditions of \cite{Troessaert:2017jcm,Henneaux:2018cst,Henneaux:2018hdj,Henneaux:2019yax} by setting $\bar{P}_{AB} =0 =\bar{P}^{rr}$ and $\bar{G}_{AB} = \bar{\gamma}_{AB}$, the unit-round sphere metric. In particular, the parity conditions \eqref{parity} generalise those of these previous references, in which $\bar h_{rr}$ and the combination $(\bar\pi^{rr}-\bar\pi)$ were already assumed to have odd parity, and the parity of $\bar G_{AB}$ agrees with the one of the unit-round sphere metric. The relaxation of the definition of asymptotic flatness considered here is inspired by the analysis at null infinity, where fluctuations of the transverse boundary metric are required to obtain Diff($S^2$) superrotations \cite{Campiglia:2014yka,Campiglia:2015yka,Compere:2018ylh}. Taking into account the falloff conditions on the lapse and the shift
\begin{equation}
    N = 1 + \mathcal O(r^{-1}),\quad N^r = \mathcal O(r^{-1}),\quad N^A = \mathcal O(r^{-2}),
\end{equation} 
one can show that the present boundary conditions are compatible with the typical behaviour of asymptotically locally flat spacetime, \textit{i.e.}, the Riemann tensor vanishes when $r\to +\infty$. 

The boundary conditions \eqref{BC} and \eqref{parity} are invariant under hypersurface deformations generated by 
\begin{equation}
\begin{split}
\xi^\bot &= r b + f + \frac{\epsilon}{r}+ \mathcal O(r^{-2}) ,  \\
\xi^{r} &= W + \frac{\epsilon^{r}}{r}+ \mathcal O(r^{-2}) ,  \\
\xi^{A} &= Y^{A} +  \frac{\epsilon^{A}}{r} + \frac{\epsilon^A_{(2)}}{r^2}+\mathcal O(r^{-3}) ,
\end{split} \label{residual gauge transfos}
\end{equation}
where $\xi^\bot$, $\xi^r$ and $\xi^A$ denote respectively infinitesimal deformations in the normal, radial and angular directions. In Eq. \eqref{residual gauge transfos}, the parameters $f,W,\epsilon,\epsilon^r,\epsilon^A,\epsilon^A_{(2)}$ are arbitrary functions on the sphere, and $b,Y^A$ are parity-odd functions by consistency with Eq. \eqref{parity}. Acting on the radial expansions in Eq. \eqref{BC} with the deformations \eqref{residual gauge transfos} in the standard way, see \textit{e.g.} \cite{Dirac:1958sc,Arnowitt:1962hi}, yields the transformations of each field defined in these expansions. 

For the BMS boundary conditions of \cite{Troessaert:2017jcm,Henneaux:2018cst,Henneaux:2018hdj,Henneaux:2019yax}, fixing the leading structure yields $\mathscr L_Y \bar\gamma_{AB} = 0$ and $\bar D_A\bar D_B b +\bar\gamma_{AB} b=0$ whose solutions define the Lorentz generators ($Y^A$ parametrises the three rotations and $b$ the three boosts). In our case, $Y^A$ and $b$ are not forced to obey these constraints and contain an infinite tower of modes, which we will identify later as the Spi superrotations. As discussed in the \hyperref[sec:Appendix]{Appendix}, one can consistently set $\bar\lambda_A = 0 = \bar h_{rA}$ by adjusting the subleading parameters $\epsilon^A,\epsilon^A_{(2)}$, which we will assume from now on.

\section{Symplectic structure}
\label{sec:Symplectic structure}

 Taking one variation of the Lagrangian density ${\bm{\mathcal L}}_H = {\mathcal L}_H\,\D t\,\D^3 x$ gives
\begin{equation}
    \delta \bm{\mathcal L}_H = \big( A^{ij}\delta g_{ij} + B_{ij}\delta \pi^{ij}\big)\,\D t\,\D^3 x - \D\bm\Theta
\end{equation}
where $A^{ij}=0$, $B_{ij}=0$ are the Hamilton (Einstein) equations and the three-form $\bm\Theta$ is the presymplectic potential obtained by keeping the boundary terms in the integrations by parts and from which the presymplectic current $\bm\omega \equiv \delta\bm\Theta$ derives. Given any Cauchy slice $\Sigma$, the presymplectic structure $\Omega=\int_\Sigma\bm\omega$ is then given by \cite{Dirac:1958sc,Arnowitt:1962hi}
\begin{equation}
    \Omega\big[\delta g_{ij},\delta \pi^{ij}\big] = \int_\Sigma \D^3 x \, \delta \pi^{ij}\wedge \delta g_{ij}. \label{Omega}
\end{equation}
For our boundary conditions \eqref{BC}, the latter is linearly divergent in $r$ due to the fluctuation of the leading fields $\bar G_{AB},\bar P^{AB}$. To renormalise these divergences, we supplement the action with some appropriate boundary term $\bm{\mathcal B}$, \textit{i.e.} $\bar{\bm{\mathcal L}}_H \equiv \bm{\mathcal L}_H - \D\bm{\mathcal B}$ \cite{Papadimitriou:2005ii,Mann:2005yr,Compere:2008us,Compere:2020lrt,Fiorucci:2020xto}. We have
\begin{equation}
    \delta \bm{\mathcal B} = \frac{\delta\bm{\mathcal B}}{\delta\phi}\delta\phi - \D\bm\theta_b,\quad \bm\omega_b = \delta\bm\theta_b \label{delta B}
\end{equation}
where $\phi$ collectively denotes relevant boundary degrees of freedom. The incorporation of the boundary term then modifies the presymplectic potential as \cite{Papadimitriou:2005ii,Mann:2005yr,Compere:2008us,Compere:2020lrt,Fiorucci:2020xto}
\begin{equation}
    \bar{\bm\Theta} = \bm\Theta - \delta\bm{\mathcal B} - \D\bm\theta_b, \quad \bar{\bm\omega} = \bm\omega + \D\bm\omega_b \label{thetabar}
\end{equation}
such that the new symplectic structure 
\begin{equation}
    \bar\Omega\big[\delta g_{ij},\delta \pi^{ij}\big] = \int_\Sigma \D^3 x \, \delta \pi^{ij}\wedge \delta g_{ij} + \oint_{\partial\Sigma} \bm\omega_b \label{renormalized omega}
\end{equation}
is modified by a surface term on the sphere at infinity. Divergent parts in $\bm{\mathcal B}$ can be fixed in such a way that $(i)$ the modified symplectic structure \eqref{renormalized omega} is finite in the $r\to+\infty$ limit, hence ensuring finiteness of the canonical generators, and $(ii)$ the variation of the action is finite on-shell. We refer to Eq. \eqref{SMBr} for the explicit expression.

\section{Charge algebra}
\label{sec:Charge algebra}
Contracting the renormalised symplectic form \eqref{renormalized omega} with a gauge transformation $\delta_\xi$ yields the following canonical generator
\begin{equation}
    \delta\int_\Sigma \D^3 x\,\big(\xi^\bot\mathcal H + \xi^i\mathcal H_i\big) = \iota_{\delta_\xi}\bar\Omega + \lim_{r\to+
    \infty}\bar{\mathcal K}_\xi .\label{contraction with diff}
\end{equation}
The surface charge $\bar{\mathcal K}_\xi$ is finite but non-integrable, \textit{i.e.} it is not a $\delta$-exact term on the phase space. Analogously to \cite{Troessaert:2017jcm,Henneaux:2018cst,Henneaux:2018hdj,Henneaux:2019yax} (see also \cite{Barnich:2007bf,Compere:2017knf,Grumiller:2019fmp,Adami:2020ugu,Alessio:2020ioh,Ruzziconi:2020wrb,Adami:2020amw,Adami:2021nnf}), we perform a field-dependent redefinition of the parameters $f\mapsto T$, $\epsilon\mapsto\tilde\epsilon$, given explicitly by Eqs. \eqref{f to T} and \eqref{redef param}, and we require that the new parameters $T,\tilde\epsilon$ are field-independent, $\delta T=0=\delta\tilde\epsilon$, so that the remaining obstruction to integrability only involves variations of the fluctuating boundary structure $\bar G_{AB},\bar P^{AB}$.

To resolve this remaining issue, we introduce a couple of boundary fields, \textit{i.e.} a symmetric tensor $C_{AB}$ and a symmetric tensor density $F^{AB}$, and modify the boundary symplectic structure as
\begin{equation}
    \bm\omega_b \mapsto \bm\omega_b + \D^2 x\big(\delta \bar P^{AB}\wedge \delta C_{AB} + \delta F^{AB}\wedge\delta \bar G_{AB}\big). \label{mod omega b integrability}
\end{equation}
Then, we are free to prescribe the transformation laws of the new fields $(C_{AB},F^{AB})$ in such a way that the modification brought by \eqref{mod omega b integrability} to the charge absorbs the remaining non-integrability, and we refer to Eqs. \eqref{delta CAB FAB pure} and \eqref{delta CAB FAB integrability} for the explicit expressions. Furthermore, the new term in Eq. \eqref{mod omega b integrability} brings a supplementary integrable piece in the charge that reads as
\begin{equation}
    \bar{\mathcal K}_{(b,Y)}^{(C,F)} = -\delta\oint_{\partial\Sigma}\D^2 x\,\Big[ b\mathcal H^{(C,F)} + Y^A\mathcal H_A^{(C,F)} \Big] . \label{add linearized gravity to the charge}
\end{equation}
Here, $\mathcal H^{(C,F)}$ and $\mathcal H_A^{(C,F)}$ are given by
\begin{equation}
    \begin{split}
        &\mathcal{H}^{(C,F)} = \sqrt{\bar{G}} \left( \bar{D}^{A} \bar{D}^{B} C_{AB} - (\triangle+1) C \right)  \\
        &\quad -\frac{2}{\sqrt{\bar{G}}} \left[ F^A_B \bar P^B_A -\bar{P} {F} -\frac{1}{4}\left( \bar P^A_B \bar P^B_A - \bar{P}^2  \right) C \right], \\
        &\mathcal{H}_{A}^{(C,F)} = -2 \bar{D}_{B} F^{B}_{\,\,\,\,\,A} + \bar P^B_C \left( \bar{D}_{A} C_B^C  - 2 \bar{D}_{B} C^C_A \right)  ,
    \end{split}
\end{equation}
which remarkably correspond to the Hamiltonian and momentum constraints for three-dimensional linearised gravity on a curved background with positive cosmological constant \cite{Bengtsson:1994vn}, the latter being described by the spatial metric $\bar G_{AB}$ and the canonical momentum $\bar P^{AB}$. 

The fact that the background has a positive cosmological constant should be expected from the asymptotic analyses of Beig, Schmidt \cite{BeigSchmidt,BeigIntegration}, Ashtekar, Romano \cite{Ashtekar:1991vb} and Friedrich \cite{FRIEDRICH199883} where, using projective geometry, spacelike infinity is a de Sitter hyperboloid. Moreover, from \eqref{add linearized gravity to the charge}, $b$ and $Y^A$ may respectively be thought as the parameters of normal and tangential deformations of the boundary theory.

Incorporating all the modifications, the final expression of the charge is integrable and found to be
\begin{widetext}
    \begin{equation}
    \boxed{
    \boxed{
        \begin{aligned}
            \bar{\mathcal{K}}_{\xi}[g_{ij},\pi^{ij}] &= \delta \oint_{\partial\Sigma} \D^2 x  \Big[ 2 T \sqrt{\bar{G}}\bar{h}_{rr}+ 2 Y^{A}  \big( \bar{G}_{AB} \bar{\pi}^{rB}_{(2)}+ \bar{h}_{AB}  \bar{\pi}^{rB} \big) +2 W \big(\bar{\pi}^{rr}-  \bar{\pi} +\bar{h}_{rr} \bar{P} - \bar{P}^{AB} \bar{h}_{AB}   \big)  \\
            & \hspace{3em} \hspace{3em}    + b  \sqrt{\bar{G}} \big(2 \bar{k}^{(2)} + \bar{k}^2 + \bar{k}^{AB}\bar{k}_{AB} -3 \bar{h}_{rr}\bar{k} \big) +\frac{2b}{\sqrt{\bar{G}}}\bar{\pi}^{rA}\bar{\pi}^{r}_{A} + \tilde{\epsilon} \sqrt{\bar{G}} + \epsilon^{r} \bar{P}\Big] + \bar{\mathcal K}_{(b,Y)}^{(C,F)}.
        \end{aligned}\label{final charge}
     }
     }
    \end{equation}
\end{widetext}
where the fields $\bar k_{AB}$ and $\bar k^{(2)}_{AB}$ appear in the radial expansion of the extrinsic curvature $K_{AB}$ of the sphere,
\begin{equation}
    K^A_B = -\frac{1}{r}\delta^A_B + \frac{1}{r^2}\bar k^A_B + \frac{1}{r^3}{\bar k^{(2)}}{}^A_B + \mathcal O(r^{-4}). \label{K}
\end{equation}
The terms collected in square brackets in Eq. \eqref{final charge} formally reproduce the result of \cite{Henneaux:2018cst,Henneaux:2018hdj,Henneaux:2019yax}, except for the last two terms in the first line, which are sensitive to the boundary momentum $\bar P^{AB}$, and the last two terms in the second line, which involve subleading symmetry parameters. As a consequence of the parity conditions \eqref{parity}, only the parity-even part of $T$ and the parity-odd part of $W$ generate improper gauge transformations, the BMS supertranslations. However, by contrast with the analysis of \cite{Henneaux:2018cst,Henneaux:2018hdj,Henneaux:2019yax}, $b$ and $Y^A$ are now arbitrary parity-odd functions that no longer obey the conformal Killing equations on the sphere. The additional modes in $b$ and $Y^A$ compared to the Lorentz generators are identified with the Spi superrotations. Finally, we also find new large gauge symmetries associated with the parity-even part of the subleading parameter $\epsilon$ and the parity-odd part of $\epsilon^r$ in the charge expression \eqref{final charge}.

Writing $\bar{\mathcal K}_\xi = \delta \bar K_\xi$ and defining the generators 
\begin{equation}
    \mathcal G_\xi[g_{ij},\pi^{ij}] \equiv \int_\Sigma \D^3 x\big(\xi^\bot\mathcal H + \xi^i\mathcal H_i\big) + \bar{K}_\xi[g_{ij},\pi^{ij}]
\end{equation}
of the asymptotic symmetries, an intricate and lengthy computation shows that
\begin{equation}
    \big\{ \mathcal G_{\xi_1},\mathcal G_{\xi_2}\big\}[g_{ij},\pi^{ij}] \approx \mathcal G_{\hat\xi}[g_{ij},\pi^{ij}] + \Lambda_{\xi_1,\xi_2}[g_{ij},\pi^{ij}]
    \label{asympSymAlge}
\end{equation}
on the constraint surface \eqref{constraints}, where $\big\{ \mathcal G_{\xi_1},\mathcal G_{\xi_2}\big\} \equiv \delta_{\xi_2}\mathcal G_{\xi_1}$ is the Poisson bracket and $\hat\xi$ is parametrised by
\begin{widetext}
    \begin{equation}
        \begin{aligned}
            \hat Y^A &= Y_1^B \partial_B Y^A_2 + \bar G^{AB}b_1 \partial_B b_2 - (1\leftrightarrow 2),\qquad \hat b = Y_1^A \partial_A b_2 - (1\leftrightarrow 2),\qquad \hat W = Y_1^A \partial_A W_2 - b_1 T_2 - (1\leftrightarrow 2) , \\
            \hat T &= Y_1^A \partial_A T_2 - 3b_1W_2 -  \bar G^{AB} \partial_A b_1 \partial_B W_2 -b_1\bar G^{AB}\bar D_A \bar D_B W_2 - (1\leftrightarrow 2), \\
            \hat{\epsilon} &= Y_{1}^{B}\partial_{B} \epsilon_2 +b_1 \bar{G}^{AB}\bar{D}_{A}\bar{D}_{B} \epsilon^{r}_{2} +4 b_1 \epsilon^{r}_{2} -b_1 \epsilon^{r}_{2}\bar{R} -(1 \leftrightarrow 2),\qquad \hat{\epsilon}^{r} = Y_{1}^{B}\partial_{B}\epsilon^{r}_2 +b_1 \epsilon_2 -(1 \leftrightarrow 2) .
        \end{aligned}\label{structure const}
    \end{equation}
\end{widetext} 
Despite the resemblance with the results of \cite{Henneaux:2018cst}, the parameters $b$ and $Y^A$ now contain an infinite tower of Spi superrotations modes. We also find two abelian ideals associted with $\epsilon$ and $\epsilon^r$, on which the superrotations are acting non-trivially. Notice that the structure constants \eqref{structure const} now depend explicitly on the particular solution through the presence of $\bar G_{AB}$. Technically, this mathematical structure is referred to as a Lie algebroid \cite{crainic2004integrability,Barnich:2017ubf,Baulieu:2023wqb}, and naturally appears when the boundary structure is allowed to fluctuate on the phase space \cite{Compere:2019bua,Compere:2020lrt,Fiorucci:2020xto}. Moreover, the asymptotic symmetry algebra \eqref{asympSymAlge} admits the following non-linear contribution:
\begin{equation}
    \Lambda_{\xi_1,\xi_2} = 2\oint_{\partial\Sigma} \D^2 x\big(b_1 T_2 - b_2 T_1\big)\bar P\bar h_{rr}. \label{cocycle}
\end{equation} 
This term would be invisible in a linear treatment of infinity. It is reminiscent of the appearance of a field-depend two-cocycle in the charge algebra at null infinity in presence of superrotations \cite{Barnich:2011mi,Barnich:2017ubf,Compere:2018ylh}. As a non-trivial consistency check, we verified explicitly that the Jacobi identity 
\begin{equation}
    \Big\{ \mathcal G_{\xi_1},  \big\{ \mathcal G_{\xi_2} , \mathcal G_{\xi_3}  
    \big\} \Big\} + \text{cyclic}(1,2,3) \approx 0
\end{equation} is satisfied. Interestingly, the presence of the non-linear contribution \eqref{cocycle} is absolutely essential for this computation to work, due to the field-dependence of the structure constants in Eq. \eqref{structure const}. 

\section{Discussion}
In this work, we proposed a consistent set of boundary conditions at spacelike infinity allowing for an enhancement of BMS symmetries with smooth Spi superrotations. An interesting observation is that these symmetries have the same parity as the Lorentz symmetries, which was shown in \cite{Henneaux:2018cst,Henneaux:2019yax,Troessaert:2017jcm,Henneaux:2018hdj,Capone:2022gme,Compere:2023qoa} to be consistent with the antipodal matching conditions advocated in \cite{Strominger:2013jfa,Kapec:2014opa,He:2014laa}. Therefore, our analysis extends the compatibility with the antipodal matching conditions for superrotation symmetries. 

We believe that the Spi superrotations identified here can be matched with the Generalised BMS symmetries \cite{Campiglia:2014yka,Campiglia:2015yka,Compere:2018ylh} uncovered at null infinity, and which manifest here in a unusual basis adapted to Hamiltonian decomposition in time and space. The precise relation between spacelike and null infinity requires using suitable coordinate systems as Beig-Schmidt \cite{BeigSchmidt,BeigIntegration} or Friedrich \cite{FRIEDRICH199883} gauges, along the lines of \cite{Compere:2011ve,Capone:2022gme,Henneaux:2018cst,Troessaert:2017jcm}, which has still to be understood with our relaxed boundary conditions. In this set-up, it would also be beneficial to identify the geometric structure associated with the boundary conditions discussed in this paper. These important questions will be addressed elsewhere. 

Besides the Spi superrotations, we also found two infinite towers of charges associated with the subleading symmetry parameters $\epsilon$ and $\epsilon^r$. These subleading symmetries are abelian and in a semi-direct sum with the Spi superrotations. This echoes some recent results obtained at null infinity by relaxing Bondi gauge fixing conditions \cite{Barnich:2010eb,Barnich:2019vzx,Ciambelli:2020ftk,Ciambelli:2020eba,Campoleoni:2022wmf,Geiller:2021vpg,Freidel:2021fxf,Geiller:2022vto,Geiller:2024amx,Mao:2024jpt}. It would be interesting to further explore this intriguing resemblance.  

Interestingly, in the process of rendering the charges integrable, the self-consistency of the Hamiltonian analysis led us to add new fields at the boundary, see Eq. \eqref{mod omega b integrability}. The latter were re-interpreted as the canonical variables for a linearised spin-two field theory at the boundary. The emergence of these boundary degrees of freedom is reminiscent of the edge mode fields \cite{Donnelly:2016auv,Freidel:2020xyx,Donnelly:2020xgu}, which have been argued in \cite{Ciambelli:2021nmv,Freidel:2021dxw} to be useful to obtain integrable charges. Our analysis constitutes an explicit realisation of this proposal. 

Finally, let us emphasise that this work combined the powerful machinery of the Hamiltonian formalism previously applied at spacelike infinity, together with covariant phase space techniques developed in parallel mostly at null infinity. In particular, this is the first time that the renormalisation of the symplectic structure is applied in the Hamiltonian formalism at spacelike infinity. This somehow concludes a long programme of finding the complete set of BMS symmetries at spacelike infinity, which started with the seminal work of Regge and Teitelboim \cite{Regge:1974zd}, and has then known a decisive turning point with the beautiful series of papers of Henneaux and Troessaert \cite{Henneaux:2018cst,Henneaux:2018hdj,Henneaux:2019yax}.

\begin{acknowledgments}
We would like to thank Oscar Fuentealba, Marc Geiller, Marc Henneaux, Alfredo P\'erez, Ricardo Troncoso and C\'eline Zwikel for useful conversations. A.F. and R.R. acknowledge the hospitality of the IFT UAM/CSIC in Madrid, where decisive parts of this work were completed. The work of A.F. is supported by the Austrian Science Fund (FWF), projects P~30822, P~32581, and P~33789. J.M. has been supported by the MCI, AEI, FEDER (UE) grants PID2021-125700NB-C21 (“Gravity, Supergravity and Superstrings”(GRASS)) and IFT Centro de Excelencia Severo Ochoa CEX2020-001007-S. R.R. is supported by the Titchmarsh Research Fellowship at the Mathematical Institute and by the Walker Early Career Fellowship in Mathematical Physics at Balliol College.

\end{acknowledgments}


\providecommand{\href}[2]{#2}\begingroup\raggedright\endgroup


\clearpage
\onecolumngrid

\appendix
\section*{Appendix}
\label{sec:Appendix}

\renewcommand{\theequation}{A.\arabic{equation}}
\setcounter{equation}{0}

In this appendix, we collect some long and intermediate technical results, including the transformation laws of the fields, the asymptotic constraints and some details on the renormalisation of the symplectic structure and the boundary spin-two theory.\\[1em]

Canonical fields are denoted by $(g_{ij},\pi^{ij})$ where $x^i$ are spacelike coordinates on Cauchy hypersurfaces. To formulate expansions at spacelike infinity, we use a spherical coordinate system, $x^i = (r,x^A)$ where $r\in\mathbb R^+_0$ is the radius and $x^A = (\theta,\varphi)$ are the usual spherical angles ($A=1,2$), in which spacelike infinity is reached in the $r\to+\infty$ limit. We denote by $\bar D_A,\bar R_{AB},\bar R$ respectively the covariant derivative, the Ricci curvature and the Ricci scalar associated with the metric $\bar G_{AB}$ on the boundary sphere, $\triangle \equiv \bar G^{AB}\bar D_A\bar D_B$ is the spatial Laplacian, and for each spherical tensor $T_{AB}$, $T \equiv \bar G^{AB}T_{AB}$ denotes the associated trace.

In bulk coordinates $x^\mu = (t,x^i)$, codimension $k$ forms ($k = 0,\dots,4$) are denoted by boldface symbols, $\bm A = A^{\mu_1\dots\mu_k}(\D^{4-k}x)_{\mu_1\dots\mu_k}$, where the codimension $k$ basis forms are defined as
\begin{equation*}
    (\D^{4-k}x)_{\mu_1\dots\mu_k} = \frac{1}{k!(4-k)!}\,\varepsilon_{\mu_1\dots\mu_k\nu_1\dots \nu_{4-k}}\,\D x^{\nu_1}\wedge\dots \wedge \D x^{\nu_{4-k}}
\end{equation*}
and $\varepsilon_{\mu_1\mu_2\mu_3\mu_4}$ is the (numerically invariant) Levi-Civita symbol in four dimensions. The convention for the exterior derivative is $\D = \D x^\sigma\partial_\sigma$ so that $\D\bm A = \partial_\sigma A^{\mu_1\dots\mu_{k-1}\sigma}(\D^{4-k+1}x)_{\mu_1\dots\mu_{k-1}}.$ Finally, bare equation numbers quoted in this document always refer to the corresponding equation in the main paper.

\subsection{Transformation laws} 

Under hypersurface deformations of parameters $\xi^\bot$ and $\xi^i = (\xi^r,\xi^A)$, the canonical fields $(g_{ij},\pi^{ij})$ transform as \cite{Dirac:1958sc,Arnowitt:1962hi}
\begin{equation}\label{transfos}
\begin{split}
    &\delta_\xi g_{ij} = \mathscr{L}_{\xi} g_{ij} +  \frac{2}{\sqrt{g}} \xi^\bot \Big(\pi_{ij} - \frac12 g_{ij} \pi \Big) , \\
    &\begin{aligned}
        \delta_\xi \pi^{ij} &= \mathscr L_\xi \pi^{ij} -  \sqrt{g}\, \xi^\bot \Big(R^{ij} - \frac{1}{2} g^{ij} R \Big) + \frac{1}{2\sqrt{g}} \xi^\bot  g^{ij}\Big(\pi_{kl} \pi^{kl} - \frac{1}{2} \pi^2 \Big) \\ 
        & \quad - \frac{2}{\sqrt{g}} \xi^\bot  \Big(\pi^{im} {\pi_{m}}^j - \frac{1}{2} \pi^{ij} \pi \Big) + \sqrt{g} \big(\xi^{\vert ij} - g^{ij} {\xi^{\vert m}}_{\vert m} \big) , 
    \end{aligned}
\end{split}
\end{equation}
where ${}_{|i}$ denotes the Levi-Civita connection associated to the induced metric $g_{ij}$ on constant time slices. Developing these transformations with the boundary conditions at spacelike infinity given by Eq. \eqref{BC} and the compatible asymptotic expansions of the parameters displayed in Eq. \eqref{residual gauge transfos}, we derive the following transformation laws for the asymptotic fields:

\begin{subequations}\label{deltas1}
    \begin{align}
        \delta_\xi \bar{h}_{rr} &= \mathscr L_Y \bar{h}_{rr} + \frac{f}{\sqrt{\bar{G}}} \big( \bar{P}^{rr} - \bar{P} \big) + \frac{b}{\sqrt{\bar{G}}}\Big[\big( \bar{\pi}^{rr} - \bar{\pi} \big) +\big(\bar{h}_{rr}\bar{P}^{rr}-\bar{h}_{AB}\bar{P}^{AB}\big) +\frac{1}{2}\big(\bar{h}_{rr}- \bar{h} \big)\big(\bar{P}^{rr}-\bar{P}  \big)\Big], \label{delta hrr} \\
        \delta_\xi \bar{\lambda}_{A} &= \mathscr L_Y \bar{\lambda}_{A}- \epsilon_{A} + \frac{b}{\sqrt{\bar{G}}} \Big[ 2 \bar{\pi}^{r}_{\,\,A} +\big(\bar{P}^{rr}  -\bar{P} \big) \bar{\lambda}_{A} +2\bar{G}_{AB}\bar{P}^{BC} \bar{\lambda}_{C} \Big] +\bar{D}_{A}W , \label{delta lambdaA} \\
        \delta_\xi \bar G_{AB} &= \mathscr L_Y \bar G_{AB} + \frac{b}{\sqrt{\bar G}}\big[2\bar P_{AB}-\bar G_{AB}(\bar P^{rr}+\bar P)\big] ,\label{delta GAB app} \\
        \delta_\xi \bar P^{AB} &= \mathscr L_Y \bar P^{AB} + \sqrt{\bar G}\big( \bar D^A\bar D^B b - \bar G^{AB}\triangle b - \bar G^{AB}b\big) - \frac{1}{2}\frac{b}{\sqrt{\bar G}}\big( \bar P^{CD}\bar P_{CD} - \bar P^2 \big). \label{delta PAB}
    \end{align}
\end{subequations}
Note that \eqref{delta GAB app} and \eqref{delta PAB} allow $\bar G_{AB}$ and $\bar P_{AB}$ to have definite parities, these functions being respectively parity-even and parity-odd, since $b,Y^A$ are both parity-odd. In particular, this implies that the choice $\sqrt{\bar G} = \sqrt{\bar\gamma}$ defines an admissible orbit of our asymptotic symmetries. Considering \eqref{delta lambdaA}, one can set $\bar\lambda_A = 0$ on the phase space by fixing $\epsilon_A = \bar D_A W + \frac{2b}{\sqrt{\bar G}}{\bar\pi^r}_{\,\,A}$.

\bigskip
With this restriction, one has
\begin{subequations}\label{deltas2}
    \begin{align}
        \delta_\xi \bar{h}^{(2)}_{rr} &= \mathscr L_Y \bar{h}^{(2)}_{rr}+\frac{f}{\sqrt{\bar{G}}}\Big[\big( \bar{\pi}^{rr} - \bar{\pi} \big) + \bar{h}_{rr}\bar{P}^{rr}+\frac{1}{2}\big(\bar{h}_{rr}-\bar{h} \big)\big(\bar{P}^{rr} - \bar{P} \big) -\bar{h}_{AB}\bar{P}^{AB}\Big] + \bar D^A W \bar D_A \bar h_{rr} -W \bar h_{rr} \nonumber \\ 
            &\quad -2 \epsilon^{r} +\frac{\epsilon}{\sqrt{\bar{G}}}\big(\bar{P}^{rr} - \bar{P} \big) +\frac{b}{\sqrt{\bar{G}}}\bigg[ \frac{1}{4}\big(\bar{P}^{rr}-\bar{P} \big) \big( \bar{h}_{AB} \bar{h}^{AB}  +\frac{1}{2} \bar h^2 -2 \bar{h}^{(2)}\big) -\frac{3}{4} \bar{h}_{rr}\bar{h}  \bar{P}^{rr}  +\frac{3}{8} \bar{h}_{rr}^{2} \bar{P}^{rr}  \nonumber \\
            &\quad + \frac{3}{2} \bar{h}_{rr}^{(2)} \bar{P}^{rr} -\frac{1}{2} \bar{\pi}^{rr} \bar{h} +\frac{3}{2} \bar{h}_{rr} \bar{\pi}^{rr} +\frac{1}{8} \bar{h}_{rr}^{2} \bar{P} -\frac{1}{2}\bar{h}_{rr}^{(2)} \bar{P} - \bar{G}_{AB}\bar{\pi}^{AB}_{(2)}+\bar{\pi}^{rr}_{(2)}-\frac{1}{2} \bar{h}_{rr} \bar{\pi} \nonumber \\
            &\quad -\bar{h}_{AB}^{(2)} \bar{P}^{AB}+ \frac{1}{2}\left(\bar{h}^{A}_{A} -\bar{h}_{rr}\right) \bar{h}_{CD}\bar{P}^{CD} +\frac{1}{4} \bar{h}_{rr} \bar{h}  \bar{P} -\bar{h}_{AB} \bar{\pi}^{AB}+\frac{1}{2} \bar{h} \bar{\pi} \bigg], \label{delta h2rr} \\
        \delta_\xi \bar{h}_{rA} &=\mathscr{L}_{Y} \bar{h}_{rA} -\bar{h}_{AB}\Big(\bar D^B W + \frac{2b}{\sqrt{\bar G}}\bar\pi^{rA}\Big) -2 \epsilon_{A}^{(2)} + \frac{2f}{\sqrt{\bar{G}}} 2 \bar{\pi}^{r}_{\,\,A} + \bar{h}_{rr} \bar{D}_{A}W + \bar{D}_{A}\epsilon^{r} \nonumber \\
            &\quad +\frac{b}{\sqrt{\bar{G}}}\Big[  \bar{h}_{rA} \big( \bar{P}^{rr} - \bar{P} \big)+ 2 \bar{h}_{AB}\bar{\pi}^{rB} + 2 \bar{G}_{AB} \bar{\pi}^{rB}_{(2)} + \bar{h}_{rr} \bar{G}_{AB} \bar{\pi}^{rB} - \bar{h} \bar{G}_{AB} \bar{\pi}^{rB}+2 \bar{G}_{AB}\bar{P}^{BC} \bar{h}_{rC} \Big],  \label{delta hrA} \\
        \delta_\xi \bar{h}_{AB} &= \mathscr{L}_{Y}\bar{h}_{AB}+\mathscr{L}_{\epsilon^C}\bar{G}_{AB}+\frac{f}{\sqrt{\bar{G}}}\big[2\bar{P}_{AB} -\bar{G}_{AB}\big(\bar{P}^{rr}+\bar{P} \big) \big]+ 2 W \bar{G}_{AB} +\frac{b}{\sqrt{\bar{G}}}\Big[  -\bar{G}_{AB} \bar{\pi} + 2 \bar{\pi}_{AB}   \nonumber \\
            &\quad - \bar{\pi}^{rr} \bar{G}_{AB}-\frac{1}{2}\bar{G}_{AB} \big(\bar{h}_{rr}-\bar{h} \big) \big(\bar{P}^{rr} - \bar{P} \big)   - \bar{h}_{rr} \bar{P}_{AB}- \bar{h}_{AB} \bar{P}^{rr} + 2 \bar{h}_{C(A}\bar{P}^{\,\,C}_{B)} \Big], \label{delta hAB} \\
        \delta_\xi \bar{\pi}^{rr} &=  \mathscr L_Y \bar{\pi}^{rr} + \mathscr L_{\epsilon^{C}} \bar{P}^{rr} -\sqrt{\bar{G}} \triangle f -\frac{f}{2\sqrt{\bar G}}\Big[  \mathcal{H}_{0} +\frac{3}{2} (\bar{P}^{rr})^2- \bar{P}^{rr} \bar{P} -\frac{1}{2} \bar{P}^2 - 4\left(\bar P_{AB}\bar P^{AB} - \bar{P}^2\right)\Big] \nonumber \\
        & \quad+ W \bar{P} +\sqrt{\bar{G}} \bar{h}_{AB} \bar{D}^{A}\bar{D}^{B}b + \frac{1}{2}\sqrt{\bar{G}} (\bar{h}_{rr} -\bar{h} ) \triangle b +\sqrt{\bar{G}} \bar{D}_{A} \bar{h}^{AB} \bar{D}_{B} b -\frac{1}{2} \sqrt{\bar{G}} \bar{D}^{A} \bar{h} \bar{D}_{A} b  \nonumber \\
        & \quad +\frac{b}{\sqrt{\bar G}}\Big[ \frac{1}{2} \big(\triangle - \bar h-\bar R - \bar P_{AB}\bar P^{AB}+5\bar G\big)\bar h_{rr} + \frac{1}{2} \left( \bar P_{AB} \bar P^{AB} - \bar{P}^2\right)\bar{h} + \bar{P} \bar P^{AB}\bar{h}_{AB} \nonumber \\
        & \quad -\big(\bar\pi+\bar\pi^{rr}-2\bar P_{AB}\bar\pi^{AB}\big) -\frac{1}{2}\sqrt{\bar G}\mathcal H_{-1}\Big], \label{delta pi rr} \\ 
        \delta_\xi \bar{\pi}^{rA} &= \mathscr L_Y \bar{\pi}^{rA} +\bar{P}  \Big(\bar D_A W + \frac{2b}{\sqrt{\bar G}}{\bar\pi^r}_{\,\,A}\Big) -\sqrt{\bar{G}} \bar{D}^{A} f -\bar{P}^{AB}\bar{D}_{B} W -2 \frac{b}{\sqrt{\bar{G}}} \bar{P}^{AB} \bar{\pi}^{r}_{\,\,B}- b \sqrt{\bar{G}} \bar{D}^{A} \bar{h}_{rr} \nonumber \\
            &\quad + \frac{1}{2} \sqrt{\bar{G}} \bar{h}^{AB} \bar{D}_{B} b + \frac{1}{2}  \sqrt{\bar{G}} b \big(\bar{D}_{B}\bar{h}^{BA} - \bar{D}^{A} \bar{h}^{B}_{B} \big) + \frac{1}{2}\frac{b}{\sqrt{\bar G}}\big(\bar P^{BC}\bar P_{BC} - \bar P^2\big), \label{delta pi rA} \\
        \delta_\xi \bar \pi^{AB} &= \mathscr{L}_{Y} \bar{\pi}^{AB} +\mathscr{L}_{\epsilon^C} \bar{P}^{AB} +  \sqrt{\bar{G}} \big(\bar{D}^{A}\bar{D}^{B} f - \bar{G}^{AB} \triangle f \big)  +\frac{1}{2} \sqrt{\bar{G}} \big( \bar{h}_{rr} - \bar{h} \big) \big( \bar{D}^{A}\bar{D}^{B} b -\bar{G}^{AB} \triangle b \big)\nonumber \\ 
            &\quad +\frac{1}{2}\sqrt{\bar{G}} b \big( \bar{D}^{A}\bar{D}^{B} \bar{h}_{rr}-  \bar{G}^{AB} \triangle \bar{h}_{rr}\big) + \frac{1}{2} \sqrt{\bar{G}}  \bar{h}^{AB} - W \bar{P}^{AB} -\frac{1}{2} \frac{f}{\sqrt{\bar{G}}}  \big(\bar{P}^{CD}\bar{P}_{CD}- \bar{P}^2 \big) \bar{G}^{AB} \nonumber \\
            &\quad -\sqrt{\bar{G}} \Big[\frac{1}{2}\bar{G}^{AB} \bar{D}^{C}\big(\bar h _{rr}+\bar{h} \big) \bar{D}_{C} b - \bar{G}^{AB} \bar{D}_{C}\bar{h}^{CD}\bar{D}_{D} b+ \bar{D}^{(A}\bar{h}^{B)C} \bar{D}_{C} b - \frac{1}{2} \bar{D}^{C}\bar{h}^{AB} \bar{D}_{C} b \Big], \label{delta piAB}
    \end{align}
\end{subequations}
where $\mathcal H_0$ and $\mathcal H_{-1}$ in \eqref{delta pi rr} represent respectively the leading and subleading orders in the radial expansion (\textit{i.e.} $\mathcal O(r^0)$ and $\mathcal O(r)$ terms) of the Hamiltonian constraint $\mathcal H$. 

\bigskip
To simplify the calculations, one can further fix $\bar h_{rA} = 0$ on the phase space. This can be done in a way that is compatible with the asymptotic conditions by fixing $\epsilon^A_{(2)}$, see \eqref{delta hrA}. Finally, \eqref{delta hrr} shows that $\bar h_{rr}$ has a definite parity, and considering \eqref{delta GAB app}, \eqref{delta PAB}, \eqref{delta hAB} and \eqref{delta pi rr}, one can show that the combination $\bar\pi^{rr}-\bar\pi+\bar h_{rr}\bar P - \bar P^{AB}\bar h_{AB}$ also has a definite parity (which turns out to be odd). Thus, the parity conditions given by Eq. \eqref{parity} are compatible with the action of the symmetries.

\subsection{Asymptotic constraints} 

For the boundary conditions given by Eq. \eqref{BC} with the additional gauge fixing $\bar\lambda_A = \bar{h}_{rA} = 0$ discussed in the previous section, the asymptotic expansion of the Hamiltonian and momentum constraints  $(\mathcal H,\mathcal H_i)$, see Eq. \eqref{constraints}, read
\begin{subequations}
\begin{align}
    \mathcal{H} &= -\sqrt{\bar{G}} \big(\bar{R}-2\big) +\frac{1}{\sqrt{\bar{G}}} \big(\bar{P}^{AB} \bar{P}_{AB} -\bar{P}^2 \big) + \frac{1}{r}\mathcal{H}_{-1}+\mathcal O(r^{-2}), \\ 
    \mathcal{H}_{r} &= -2 \big(\bar{P}^{rr} - \bar{P} \big)+\frac{1}{r}\big(-2 \bar{D}_{A}\bar{\pi}^{rA} - \bar{h}_{rr} \bar{P}^{rr} + \bar{h}_{AB}\bar{P}^{AB} + 2 \bar{\pi} \big) + \mathcal O(r^{-2}), \\
    \mathcal{H}_{A} &= -r \big( 2 \bar{D}_{B}\bar{P}^{B}_{\,\,A} \big) +\Big[\bar{P} \bar{D}_{A}\bar{h}_{rr} -2 \bar{\pi}^{r}_{\,\,A} - 2 \bar{D}_{B}\bar{\pi}^{B}_{\,\, A}+\bar{D}_{A}\bar{h}_{BC} \bar{P}^{BC} - 2 \bar{D}_{C}\big(\bar{h}_{AB} \bar{P}^{BC} \big)  \Big] + \mathcal O(r^{-1}).
\end{align}
\end{subequations}
In particular, the equality $\bar P = \bar P^{rr}$, imposed by $\mathcal H_r \approx 0$ at leading order, has been used throughout the paper.

\subsection{Symplectic structure \& infinitesimal charges} 

An arbitrary variation of the Lagrangian density $\bm{\mathcal L}_H = \mathcal L_H(\D^4 x)$ defines the Hamiltonian presymplectic potential three-form $\bm\Theta$ through Eq. (7). The latter, integrated over any Cauchy slice $\Sigma$ in spacetime, yields the usual quantity
\begin{equation}
    \int_\Sigma \bm\Theta = \int_\Sigma \D^3 x \big(\pi^{ij}\delta h_{ij}\big) ,
\end{equation}
where we write $(\D^3 x)_t \equiv \D^3 x$ to ease the notation. For the boundary conditions summarised by Eq. (3), this integral is found to be \textit{radially} divergent, \textit{i.e.}
\begin{equation}
\begin{split}
    \int_\Sigma \bm\Theta &= \int_\Sigma \D^3 x \Big[ r \big(\bar{P}^{AB} \delta \bar{G}_{AB}\big) + \big( \bar{P}^{AB} \delta \bar{h}_{AB} + \bar{\pi}^{AB} \delta \bar{G}_{AB} + \bar{P}^{rr} \delta \bar{h}_{rr} \big) \\
&\quad\quad +\frac{1}{r} \big( \bar{P}^{AB} \delta \bar{h}^{(2)}_{AB} +  \bar{\pi}^{AB} \delta \bar{h}_{AB}+\bar{\pi}_{(2)}^{AB} \delta \bar{G}_{AB} + \bar{P}^{rr} \delta \bar{h}^{(2)}_{rr} + \bar{\pi}^{rr} \delta \bar{h}_{rr}  \big) \Big] + \mathcal O(r^{0}).
\end{split} \label{div theta}
\end{equation}

\bigskip
Due to the parity conditions imposed in Eq. (4), the first term of \eqref{div theta} vanishes and the presymplectic potential is at most linearly divergent. The divergences are subtracted by introducing a boundary Lagrangian, \textit{i.e.} a codimension one form $\bm{\mathcal B}$ as
\begin{equation}
	\bar{\bm{\mathcal L}}_H = \bm{\mathcal L}_H - \D\bm{\mathcal B},\qquad \bm{\mathcal B} = \mathcal B^r(\D^3 x)_r + \mathcal B^t(\D^3 x)_t + \mathcal B^A(\D^3 x)_A, \label{bndlag}
\end{equation}
and taking advantage of the ambiguities in defining the presymplectic potential as in Eq. (11), to change it as
\begin{equation}
	\bar{\bm\Theta} = \bm\Theta - \delta\bm{\mathcal B} - \D\bm\theta_b, \qquad \bar{\bm\omega} = \bm\omega + \D\bm\omega_b \label{thetabar}
\end{equation}
using $\bm{\mathcal B}$ and the associated corner presymplectic potential $\bm\theta_b$ \cite{Papadimitriou:2005ii,Mann:2005yr,Compere:2008us,Compere:2020lrt,Fiorucci:2020xto}. Since the boundary Lagrangian is a top-form on the timelike boundary $\mathcal M_\infty = \{r \to+\infty\}$, one can freely set $\mathcal B^t = 0$. Moreover, the angular components $\mathcal B^A$ do not matter as the final result is integrated on the sphere at infinity, $\partial\Sigma\subset\mathcal M_\infty$. Integrating \eqref{thetabar} on the constant time slice $\Sigma$ then yields
\begin{equation}
    \int_\Sigma  \D^3 x \,\bar\Theta^t = \int_\Sigma \D^3 x\big(\Theta^t - \partial_r\theta^{tr}_b\big) = \int_\Sigma  \D^3 x \,\Theta^t - \oint_{\partial\Sigma} \D^2 x \, \theta_b^{tr} \label{theta bar t}
\end{equation}
where $\D^2 x \equiv (\D^2 x)_{tr}$, again to ease the notation, and $\theta_b^{tr}$ is taken to be the boundary term resulting from the variation of the boundary Lagrangian $\bm{\mathcal B}$ after performing the integration by parts on time derivatives, see the $r$-component of Eq. (11), \textit{i.e.} 
\begin{equation}
	\delta \mathcal B^r = \frac{\delta \mathcal B^r}{\delta\phi}\delta\phi - \partial_t \theta^{tr}_b - \partial_A \theta^{Ar}_b,
\end{equation}
where $\phi$ denotes the collection of relevant boundary fields. This observation leads us to set $\mathcal B^r$ as
\begin{equation}
    \mathcal B^r =  r\big( \bar{P}^{AB}\dot{\bar{h}}_{AB} + \bar{\pi}^{AB} \dot{\bar{G}}_{AB} + \bar{P}^{rr} \dot{\bar{h}}_{rr} \big) + \ln r \big( \bar{P}^A_B \dot{\bar{h}}^{(2)B}_{A} +  \bar{\pi}^A_B \dot{\bar{h}}_A^B +\bar{\pi}_{(2)}^{AB} \dot{\bar{G}}_{AB} + \bar{P}^{rr} \dot{\bar{h}}^{(2)}_{rr} + \bar{\pi}^{rr} \dot{\bar{h}}_{rr}  \big)
    \label{SMBr}
\end{equation}
to remove the divergences. Note that the above requirement of charge finiteness does not completely fix \eqref{Br}, as putative additional terms without time derivatives would not have any impact on the charges, and can then be incorporated for free. It is also worthwhile to remark that this framework is fairly general. It has for instance been applied in the context of electromagnetism in \cite{Henneaux:2018gfi} to make the generators integrable thanks to the introduction of a new boundary degree of freedom.

\bigskip
The resulting symplectic structure 
\begin{equation}
    \bar\Omega\big[\delta g_{ij},\delta \pi^{ij}\big] = \int_\Sigma \D^3 x \, \big(\delta \pi^{ij}\wedge \delta g_{ij}\big) + \oint_{\partial\Sigma} \bm\omega_b
\end{equation}
is finite and contracting it with an asymptotic symmetry generator $\xi$, see Eq. (12), we obtain the infinitesimal charges
\begin{equation}
    \begin{split}
        \bar{\mathcal K}_\xi &= \oint \D^2 x \bigg\{ 2 Y^{A} \delta \big( \bar{G}_{AB} \bar{\pi}^{rB}_{(2)}+ \bar{h}_{AB}  \bar{\pi}^{rB} \big) + 2W \delta \big( \bar{\pi}^{rr}-  \bar{\pi} +\bar{h}_{rr} \bar{P} - \bar{P}^{AB} \bar{h}_{AB} \big) \\
        &\qquad + 2\Big(f + \frac{1}{2}b\big(\bar h + 3\bar h_{rr}\big)\Big)\delta\big(\sqrt{\bar G}\bar h_{rr}\big) + b\, \delta \Big[ \sqrt{\bar{G}} \big(2 \bar{k}^{(2)} + \bar{k}^2 + \bar{k}^{A}_{B}\bar{k}^{B}_{A} -3 \bar{h}_{rr}\bar{k} \big) +\frac{2}{\sqrt{\bar{G}}}\bar{\pi}^{rA}\bar{\pi}^{r}_{A}\Big] \\
        &\qquad +\big( W\bar{h}_{AB}+W\bar{h}_{rr} \bar{G}_{AB}+2\epsilon^{r}  \bar{G}_{AB}\big)\delta \bar{P}^{AB} \\
        &\qquad - \left[ 4\epsilon + f\left(\bar h_{rr} + \bar h\right) -\frac{3b}{2}\left(\bar h^2_{rr} + \bar h_{rr}\bar h + \frac{1}{2}\bar h^2 \right) + 2b \left(\bar h^{(2)} + \bar h_{rr}^{(2)} - \frac{1}{2}\bar h^A_B\bar h^B_A - \frac{1}{\bar G}\bar\pi^{rA}\bar\pi^r{}_A\right)\right]\delta \sqrt{\bar G} \\
        &\qquad +\Big[ b\sqrt{\bar{G}} \Big( \bar{h}^{(2)AB} -\frac{1}{4} \bar{h}_{rr} \bar{h}^{AB} -\frac{1}{2} \bar{h}^{A}_{\,\,C} \bar{h}^{CB} +\frac{1}{4} \bar{h} \bar{h}^{AB}\Big) +\frac{2b}{\sqrt{\bar{G}}}\bar{\pi}^{rA}\bar{\pi}^{rB} +\frac{1}{2} \Big(f + \frac{1}{2}b\big(\bar h + 3\bar h_{rr}\big)\Big) \sqrt{\bar{G}}\bar{h}^{AB} \\
        &\qquad + \epsilon^{r} \bar{P}^{AB} + W(\bar{h}_{rr} \bar{P}^{AB} - \bar{\pi}^{AB})+2 \bar{\pi}^{rA}\bar{D}^{B}W   \Big] \delta \bar{G}_{AB}\bigg\}.
    \end{split} \label{non int charge}
\end{equation}
In this formula, $\bar k_{AB}$ and $k_{AB}^{(2)}$ appear in the radial expansion of the extrinsic curvature of two-spheres, see Eq. (17), and read explicitly as
\begin{equation}
    \bar k_{AB} = \frac{1}{2}\big(\bar h_{AB} + \bar h_{rr} \bar G_{AB}\big),\qquad
        \bar k^{(2)}_{AB} = \bar h^{(2)}_{AB} - \frac{1}{2}\left(\bar h_A{}^C  -\frac{1}{2}{\delta_A}^C\bar h_{rr}\right)\bar h_{CB} + \frac{1}{2}\left( 
\bar h^{(2)}_{rr} - \frac{3}{4}\bar h^2_{rr} \right)\bar G_{AB}.
\end{equation}
The first line of \eqref{non int charge} is readily integrable assuming $\delta Y^A = 0 = \delta W$. The third and sixth terms, involving $\delta \bar h_{rr}$ and $\delta\sqrt{\bar G}$ respectively, can be made directly integrable by redefining the parameters $f$ and $\epsilon$ as 
\begin{subequations}
    \begin{align}
        T &\equiv  f + \frac{b}{2}\big(\bar h + 3\bar h_{rr}\big), \label{f to T} \\
        \tilde\epsilon &\equiv -4\epsilon - T\left(\bar h_{rr} + \bar h\right) + \frac{3b}{2}\left(\bar h^2_{rr} + \bar h_{rr}\bar h + \frac{1}{2}\bar h^2 \right) - 2b \left(\bar h^{(2)} + \bar h_{rr}^{(2)} - \frac{1}{2}\bar h^A_B\bar h^B_A - \frac{1}{\bar G}\bar\pi^{rA}\bar\pi^r{}_A\right),
        \label{redef param}
    \end{align}
\end{subequations}
and assuming $\delta T = 0 = \delta\tilde\epsilon$. Note that the redefinition \eqref{f to T} has already been considered in \cite{Henneaux:2018cst,Henneaux:2018hdj,Henneaux:2019yax} in order to make the supertranslation charge integrable.

\subsection{Variational principle}
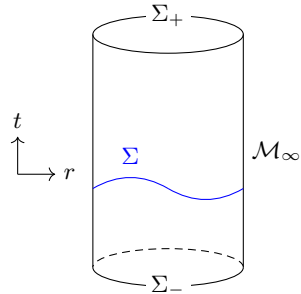
\begin{figure}[ht!]
	\centering
	\begin{tikzpicture}[scale=0.5]
		\def\dx{.2};
		\def\dy{.4};
		\def\shiftx{6.6};
		\def\shifty{-7};
		\coordinate (A) at (-2,-7+\dy);
		\coordinate (B) at (2,-7+\dy);
		\coordinate (C) at (2,-\dy);
		\coordinate (D) at (-2,-\dy);
		\coordinate (E) at (-2+\shiftx,\dy+\shifty);
		\coordinate (F) at (2+\shiftx,\dy+\shifty);
		\coordinate (G) at (2+\shiftx,7-\dy+\shifty);
		\coordinate (H) at (-2+\shiftx,7-\dy+\shifty);
		\def\decal{0.1};
		\draw (E) -- (H);
		\draw (F) -- (G);
		\draw ($(F)!0.5!(G)$) node[right]{$\mathcal M_\infty$};
		\fill [white] (F) arc[x radius=2, y radius=0.5, start angle=0, end angle=360];
		\draw (F) arc[x radius=2, y radius=0.5, start angle=0, end angle=-180];
		\draw [densely dashed] (F) arc[x radius=2, y radius=0.5, start angle=0, end angle=180];
		\draw (G) arc[x radius=2, y radius=0.5, start angle=0, end angle=360];
		\path [blue] (-2+\shiftx,2.5+\shifty) edge[bend left=30] (0+\shiftx,2.5+\shifty) -- (0+\shiftx,2.5+\shifty) edge[bend right=30] 	(2+\shiftx,2.5+\shifty);
		\draw (-1+\shiftx,3+\shifty) node[above,blue,inner sep=2pt]{$\Sigma$};
		\coordinate (C1) at ($(F)!0.9!(G)$);
		\coordinate (C2) at ($(F)!0.5!(G)$);
		\coordinate (C3) at ($(F)!0.1!(G)$);
		\coordinate (D1) at ($(C1)+(3*\decal,0)$);
		\coordinate (D2) at ($(C2)+(2.5+3*\decal,0)$);
		\coordinate (D3) at ($(C3)+(3*\decal,0)$);
		\draw ($(G)!0.5!(H)+(0,0.5)$) node[fill=white,inner sep=2pt]{$\Sigma_+$};
		\draw ($(E)!0.5!(F)-(0,0.5)$) node[fill=white,inner sep=2pt]{$\Sigma_-$};
		\coordinate (rep) at ($(E)!0.4!(H)-(2,0)$);
		\draw[->] (rep) -- ($(rep)+(1,0)$)node[right]{$r$};
		\draw[->] (rep) -- ($(rep)+(0,1)$)node[above]{$t$};
	\end{tikzpicture}
\caption{\textit{Finite volume $\mathcal V$ to define the variational principle}}
\label{fig:var}
\end{figure}

In this part, we show that the variation of the action can be made finite on-shell using the procedure introduced in Section III of \cite{Fiorucci:2024ndw} and re-discussed in Section \ref{sec:sympls} above. We formulate the variational principle
\begin{equation}
	S = \int\D t\left[\int\D^3 x \left(\pi^{ij}\dot g_{ij} - N\mathcal H - N^i\mathcal H_i\right) - B_\infty\right],\quad B_\infty \equiv \lim_{r\to+\infty} \oint (\D^2 x)\mathcal B^r,
\end{equation}
including the boundary term $B_\infty$, in a finite volume $\mathcal V$ depicted by Figure \ref{fig:var}. The latter is limited in the past and in the future by (portions of) Cauchy slices $\Sigma_-$ and $\Sigma_+$ and its spacelike boundary $\mathcal M_\infty$ is located at fixed $r\to+\infty$. As usual, $(N,N^i)$ denote the lapse and shift functions.

Defining the new Lagrangian as \eqref{bndlag} and taking \eqref{thetabar} into account, one gets on-shell
\begin{equation}
\begin{split}
	\delta S &= \int_{\mathcal V} \delta \bar{\bm{\mathcal L}}_H = -\int_{\Sigma_+}(\D^3 x)_t\,\bar\Theta_H^t + \int_{\Sigma_-}(\D^3 x)_t\,\bar\Theta_H^t - \int_{\mathcal M_\infty}(\D^3 x)_r\,\bar\Theta_H^r \\
	&= -\int_{\Sigma_+ - \Sigma_-}(\D^3 x)_t\big(\pi^{ij}\delta g_{ij} - \delta\mathcal B^t - \partial_r\theta^{tr}_b\big)  - \int_{\mathcal M_\infty}(\D^3 x)_r\big(\Theta_H^r - \delta\mathcal B^r - \partial_t\theta^{rt}_b\big) \\
	&= -\int_{\Sigma_+ - \Sigma_-}(\D^3 x)_t\big(\pi^{ij}\delta g_{ij} - \delta\mathcal B^t\big) - \int_{\mathcal M_\infty}(\D^3 x)_r\big(\Theta_H^r - \delta\mathcal B^r\big) + \int_{\partial(\Sigma_+ - \Sigma_-)}\D^2 \Omega \big(\theta^{tr}_b + \theta^{rt}_b \big)
\end{split} \label{deltaS}
\end{equation}
where total derivatives on the boundary sphere vanish upon integration and we used the shorthand notation
\begin{equation}
    \int_{\Sigma_+ - \Sigma_-}(\D^3 x)_t \, f \equiv \int_{\Sigma_+} (\D^3 x)_t \, f - \int_{\Sigma_-} (\D^3 x)_t \, f.
\end{equation}
The last term in \eqref{deltaS} trivially cancels out, and the first term also disappears as implied by Dirichlet boundary conditions in time, \textit{i.e.} $\delta g_{ij}(t\to\pm\infty) = 0$, and our choice $\mathcal B^t = 0$. We then conclude that the variational principle is well-defined (or ``the total Hamiltonian is differentiable'') if and only if the boundary term $B_\infty$ absorbs all the residues from the spatial integrations by parts, \textit{i.e.} $\delta \mathcal B^r = \Theta^r_H$, which is the seminal result of \cite{Regge:1974zd}.

\bigskip
For our choice of boundary conditions, we find explicitly
\begin{equation}
    \Theta^r_H = - 2r \delta\sqrt{\bar G} + \mathcal O(r^0).
\end{equation}
As mentioned in the previous section, the $r$-component $\mathcal B^r$ of the boundary Lagrangian $\bm{\mathcal B}$, fixed at the moment as Eq. \eqref{Br}, is only determined up to some terms that do not involve time derivatives, as they do not affect the computation of the charges. We can use this freedom to make the variation of the action finite on-shell. Indeed, taking instead
\begin{equation}
    \begin{split}
        \mathcal B^r &\approx -2r\sqrt{\bar G} + r\bar P^{AB}\big(\dot{\bar h}_{AB} - 2 \bar P_{AB} + 2 \bar P\bar G_{AB}\big) + r\big( \bar{\pi}^{AB} \dot{\bar{G}}_{AB} + \bar{P}^{rr} \dot{\bar{h}}_{rr} \big) \\
        &\quad +\frac{\ln r}{\sqrt{\bar  G}}\, \bar P^{AB} \Big[\dot{\bar h}_{AB}^{(2)} + \bar P_{AB}(\bar h_{rr}-\bar h) - 2\bar D_{(A}\bar\pi^r{}_{B)} - 2\bar\pi_{AB} - \bar G_{AB}\big(\bar\pi + \bar\pi^{rr} - \bar h_{CD}\bar P^{CD} + \bar h\bar P\big) \Big] \\
        &\quad + \ln r\, \bar \pi^{AB}\big( 
\dot{\bar h}_{AB} - 2 \bar P_{AB} + 2 \bar P\bar G_{AB} \big) + \ln r \big( \bar{\pi}_{(2)}^{AB} \dot{\bar{G}}_{AB} + \bar{\pi}^{rr} \dot{\bar{h}}_{rr} \big) \\
&\quad + \frac{\ln r}{\sqrt{\bar G}} \bar P^{rr}\big(\dot{\bar h}^{(2)}_{rr} - \bar\pi^{rr} + \bar\pi - \bar h_{rr}\bar P^{rr} + \bar h_{AB}\bar P^{AB} \big)
    \end{split}
\end{equation}
while neglecting terms vanishing on the constraint surface, ensures that the remaining part of \eqref{deltaS} vanishes on-shell. In particular, the terms in the brackets vanish separately on-shell.

\subsection{Elements from the boundary graviton theory} 
New boundary fields $C_{AB},F^{AB}$ have been introduced in order to integrate the charges \eqref{non int charge}. They are shown to correspond to a boundary graviton field and its canonical momentum. The Hamiltonian action for a linearised spin-two field propagating over a two-dimensional curved background, described by a pair $(\bar G_{AB},\bar P^{AB})$ of canonical fields, is written
\begin{equation}
    S = \int\D t\, \D^2 x \big(F^{AB}\dot{\bar G}_{AB} + \bar P^{AB}\dot C_{AB} - n\mathcal H^{(C,F)} - n^A \mathcal H_A^{(C,F)}\big),
\end{equation}
where $n,n^A$ represent the lapse and shift functions and the Hamiltonian and momentum constraints are given by \cite{Bengtsson:1994vn} 
\begin{subequations}
    \label{constraints spintwo}
    \begin{align}
        \mathcal{H}^{(C,F)} =& \sqrt{\bar{G}} \left( \bar{D}^{A} \bar{D}^{B} C_{AB} - \triangle C - 2\Lambda C \right)  -\frac{2}{\sqrt{\bar{G}}} \left( F^{AB} \bar P_{AB} -\bar{P} {F}  \right)  -\frac{1}{2\sqrt{\bar{G}}} \left( \bar P^{AB} \bar P_{AB} - \bar{P}^2  \right) C , \label{H1} \\
        \mathcal{H}_{A}^{(C,F)} =& -2 \bar{D}_{B} F^{B}_{\,\,\,\,\,A} + \bar P^{BC} \left( \bar{D}_{A} C_{BC}  - 2 \bar{D}_{B} C_{CA} \right)  .
\end{align}
\end{subequations}
In our case, $\Lambda = 1/2$ and the background is diffeomorphic to de Sitter spacetime. This observation is consistent with Lagrangian analyses that usually fix in the so-called Beig-Schimdt gauge around spacelike infinity \cite{BeigSchmidt,BeigIntegration,Compere:2011ve,Compere:2017knf} (see also \cite{Ashtekar:1991vb} for the covariant approach). 

Under constant time hypersurface deformations parameterised by $(b,Y^A)$, the canonical fields transform as 
\begin{equation}
    \delta_{b,Y}^{\textrm{\tiny bnd}} C_{AB} = \oint_{\partial\Sigma}\D^2 x\,\left\lbrace C_{AB}, b\mathcal H^{(C,F)}+Y^A\mathcal H_A^{(C,F)} \right\rbrace,\quad \delta_{b,Y}^{\textrm{\tiny bnd}} F^{AB} = \oint_{\partial\Sigma}\D^2 x\,\left\lbrace F^{AB}, b\mathcal H^{(C,F)}+Y^A\mathcal H_A^{(C,F)} \right\rbrace
\end{equation}
through the Poisson bracket. A lengthy computation then yields the following explicit transformation laws:
\begin{subequations}\label{delta CAB FAB pure}
\begin{align}
    &\delta_{b,Y}^{\textrm{\tiny bnd}} C_{AB} = \mathscr{L}_{Y} C_{AB} + \frac{2b}{\sqrt{\bar{G}}}\big(F_{AB} - \bar G_{AB} F \big) + \frac{b}{\sqrt{\bar{G}}} C \big( \bar P_{AB} - \bar{G}_{AB}\bar{P} \big) , \label{delta CAB pure} \\
    &\begin{aligned}
        \delta_{b,Y}^{\textrm{\tiny bnd}} F^{AB}  &= \mathscr L_Y F^{AB}-\frac{1}{2} \sqrt{\bar{G}} {C} \big(\bar{D}^{A} \bar{D}^{B}b-\bar{G}^{AB} \triangle b \big)    -\frac{b}{\sqrt{\bar{G}}}\bar{G}^{AB}\big( \bar P_{CD} F^{CD}  - \bar{P} {F} \big) \\
        &\quad + b \sqrt{\bar{G}}  C^{AB} - \frac{1}{2}\sqrt{\bar{G}} \bar{D}_{C}b\Big[  2\bar{D}^{(A} C^{B)C} -\bar{D}^{C} C^{AB}- \bar{G}^{AB} \big( 2\bar{D}_{D}C^{CD} - \bar{D}^{C} {C}  \big)\Big]     \\
        &\quad +\frac{b}{2\sqrt{\bar{G}}}\left(C^{AB} -\frac{1}{2}\bar{G}^{AB} C\right)\big(\bar P_{CD}\bar P^{CD} - \bar{P}^2\big) + b \sqrt{\bar{G}}\left(C^{AB} - \frac{1}{2} {C} \bar{G}^{AB}\right). 
    \end{aligned}
    \label{delta FAB pure}  
\end{align}
\end{subequations}
Finally, to make the charge \eqref{non int charge} integrable, the latter have to be supplemented by bulk contributions, so that the complete variations of the boundary fields $(C_{AB},F^{AB})$ take the following form:
\begin{subequations} \label{delta CAB FAB integrability}
    \begin{align}
        \delta_\xi C_{AB} &=  \delta_{b,Y}^{\textrm{\tiny bnd}} C_{AB} + W\big(\bar{h}_{AB}+\bar{h}_{rr} \bar{G}_{AB}\big)+\epsilon^{r}  \bar{G}_{AB}, \\
        \delta_\xi F^{AB} &=  \delta_{b,Y}^{\textrm{\tiny bnd}} F^{AB} + W\big(\bar{\pi}^{AB}- \bar{h}_{rr} \bar{P}^{AB} \big) - 2 \bar{\pi}^{r(A}\bar{D}^{B)}W -\frac{1}{2} T \sqrt{\bar{G}}\bar{h}^{AB}  -\frac{2b}{\sqrt{\bar{G}}}\bar{\pi}^{rA}\bar{\pi}^{rB} \nonumber \\
        &\quad - b\sqrt{\bar{G}} \Big( \bar{h}^{(2)AB} -\frac{1}{4} \left(\bar{h}_{rr} + \bar{h}\right)\bar{h}^{AB} \Big) .
\end{align}
\end{subequations}

The above transformation laws conclude this technical appendix. ---

\end{document}